\documentclass[aps,prc,twocolumn,showpacs,showkeys,preprintnumbers,amsmath,amssymb,a4paper]{revtex4-1}

\usepackage{amssymb}
\usepackage{graphics}
\usepackage{psfrag}
\usepackage{epsfig}
\def\bea{\begin{eqnarray}}
\def\eea{\end{eqnarray}}
\def\beq{\begin{equation}}
\def\eeq{\end{equation}}

\newcommand{\eebar}{{e^+e^-}}
\newcommand{\pbarp}{{\bar p p}}
\newcommand{\ppbar}{{\bar p p}}
\newcommand{\nbarn}{{\bar N  N}}
\newcommand{\nnbar}{{\bar N  N}}

\newcommand{\lcbarlc}{\bar{\Lambda}_c^- {\Lambda}_c^+}
\newcommand{\ddbar}{D\bar{D}}
\newcommand{\ddbarc}{D^+D^-}

\newcommand{\kkbar}{\bar{K}{K}}
\begin{document}
\title{The $\psi$(3770) resonance and its production in $\bar pp \to D \bar D$}
\author{J. Haidenbauer$^1$ and G. Krein$^2$}

\affiliation{
$^1$Institute for Advanced Simulation and J\"ulich Center for Hadron
Physics, Forschungszentrum J\"ulich, D-52425 J\"ulich, Germany \\
$^2$Instituto de F\'{\i}sica Te\'{o}rica, Universidade Estadual
Paulista,
Rua Dr. Bento Teobaldo Ferraz, 271 - 01140-070 S\~{a}o Paulo, SP, Brazil
}

\begin{abstract}
The production of a $D\bar D$ meson-pair in antiproton-proton ($\bar p p$) 
annihilation close to the production threshold is investigated, with special 
emphasis on the role played by the $\psi$(3770) resonance. The
study is performed in a meson-baryon model where the elementary 
charm production process is described by baryon exchange. Effects of the 
interactions in the initial and final states are taken into account 
rigorously, where the latter involves also those due to the $\psi$(3770).
The predictions for the $D\bar D$ production cross section are in the 
range of $30$ -- $250$ nb, the contribution from the $\psi$(3770) 
resonance itself amounts to roughly $20$ -- $80$ nb.
\end{abstract}


\maketitle

\section{Introduction}
\label{intro}

In a recent publication~\cite{HK14} we have presented predictions for the charm-production 
reaction $\pbarp \rightarrow \ddbar$ close to the threshold based on a 
model where the elementary charm production process is described by 
baryon exchange and also in the constituent quark model.
The cross section was found to be in the order of $10$ -- $100$ nb and it 
turned out to be comparable to those predicted by other model calculations in 
the literature
\cite{Goritschnig13,Mannel12,Titov:2008yf,Kerbikov,Kaidalov:1994mda,Kroll:1988cd}.
 
The results in Ref.~\cite{HK14} suggested that the reaction 
$\ppbar \to \ddbar$ takes place predominantly in the $s$ wave, at
least for excess energies below 100 MeV. However, there is a 
well-established $p$-wave resonance, the $\psi(3770)$ ($J^{PC}=1^{--}$) 
which is seen as a pronounced structure in $e^+ e^- \to \ddbar$ \cite{Ablikim,Anashin},
for example, and which is located at only around 35~MeV above the $\ddbar$ threshold. 
The resonance decays almost exclusively (i.e. to 93$^{+8}_{-9}$ \%) into $\ddbar$ \cite{PDG}. 
This resonance was not included in our previous study~\cite{HK14}. Given its apparent  
prominence in the $\ddbar$ channel, the impact of the $\psi(3770)$ on the $\ppbar \to \ddbar$ 
cross section clearly should be explored. In particular, should it turn out that 
its contribution is rather large, then it would be certainly interesting to examine 
the energy range in question in pertinent experiments, which could be performed by the
PANDA Collaboration \cite{PANDA,Wiedner,Prencipe2014} at the future FAIR facility in 
Darmstadt. 

In the present study we consider the effect of the $\psi(3770)$ on the 
$\ppbar \to \ddbar$ cross section. The work complements our results presented 
in Ref.~\cite{HK14} and builds on the J\"ulich meson-baryon model for 
the reaction $\pbarp \rightarrow \kkbar$ \cite{Mull} where 
the extension of the model from the strangeness to the charm
sector follows a strategy similar to other studies by us on the $DN$ 
and ${\bar D}N$ interactions~\cite{Haidenbauer2007,Haidenbauer2008,Hai10}, 
and on the reaction $\pbarp \rightarrow \lcbarlc$~\cite{HK10},
namely by assuming as a working hypothesis SU(4) symmetry constraints. 

\section{The model}
\label{sec:1}
The framework in which we treat the charm-production reaction 
$\pbarp \rightarrow \ddbar$ is described in detail in Ref.~\cite{HK14}. 
Here we summarize only the principal features. 
The reaction amplitude for $\pbarp \rightarrow \ddbar$ is 
obtained within the distorted-wave Born approximation. Effects of the 
initial as well as of the final-state interactions, which play an important 
role for energies near the production threshold \cite{Haidenbauer:1991kt,Kohno},
are taken into account rigorously. The employed $\nnbar$ and $\ddbar$ amplitudes 
are solutions of Lippmann-Schwinger type scattering equations based on 
corresponding ($\nnbar$ and $\ddbar$) interaction potentials. 
 
The microscopic charm-production process itself is described by 
baryon exchange ($\Lambda_c$, $\Sigma_c$), in close analogy to an 
investigation of the strangeness production reaction 
$\pbarp \rightarrow \kkbar$ by the J\"ulich group \cite{Mull}. 
Specifically, the transition potential is derived from the corresponding 
transition in the $\bar KK$ case under the assumption of SU(4) symmetry,
see Ref.~\cite{HK14} for details. 

Because of the known sensitivity of the results for the cross sections on 
the initial $\pbarp$ interaction we examined its effect by considering  
several variants of the $\nbarn$ potential. Details of those potentials
can be found in Refs.~\cite{HK10,HK14}. Here we just want to mention that
they differ primarily in the elastic part where we consider variations
from keeping only the longest ranged contribution (one-pion exchange) to 
taking a full G-parity transformed $NN$ interaction as
done in \cite{Haidenbauer:1991kt}. 
All those models reproduce the total $\pbarp$ cross section in the relevant
energy range and, in general, describe also data on
integrated elastic and charge-exchange cross sections and even
$\pbarp$ differential cross sections, cf. Refs.~\cite{HK10,HK14}.

\begin{figure}[t]
\vspace*{+1mm}
\centerline{\hspace*{3mm}
\psfig{file=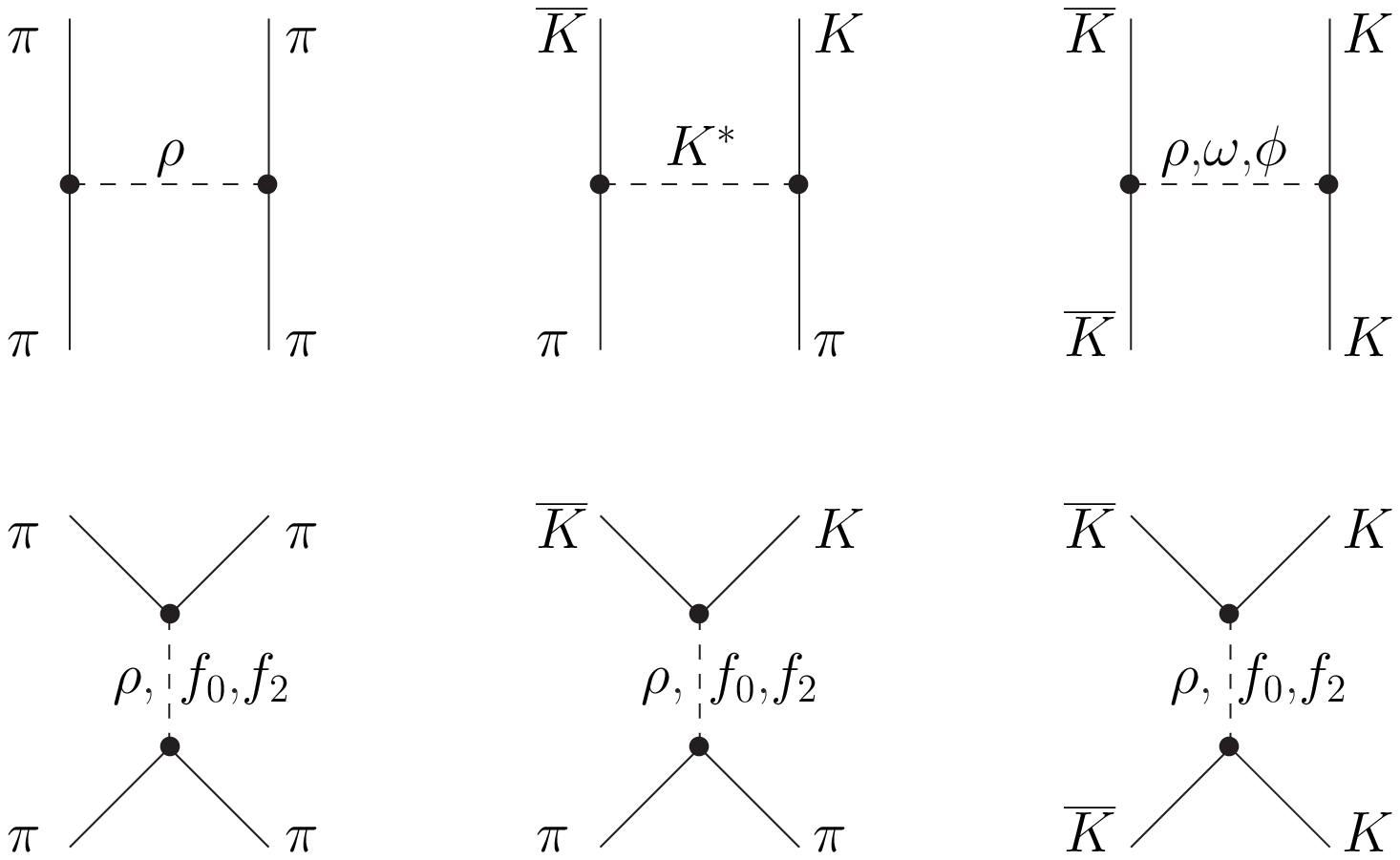,width=11.0cm,height=13.0cm}}
\vspace*{-7.5cm}
\caption{Diagrams included in the J\"ulich 
$\pi\pi- K\overline{K}$ potential \cite{Janssen95}. 
}
\label{Diag1}
\end{figure}

\begin{figure}[t]
\vspace*{+1mm}
\centerline{\hspace*{25mm}
\psfig{file=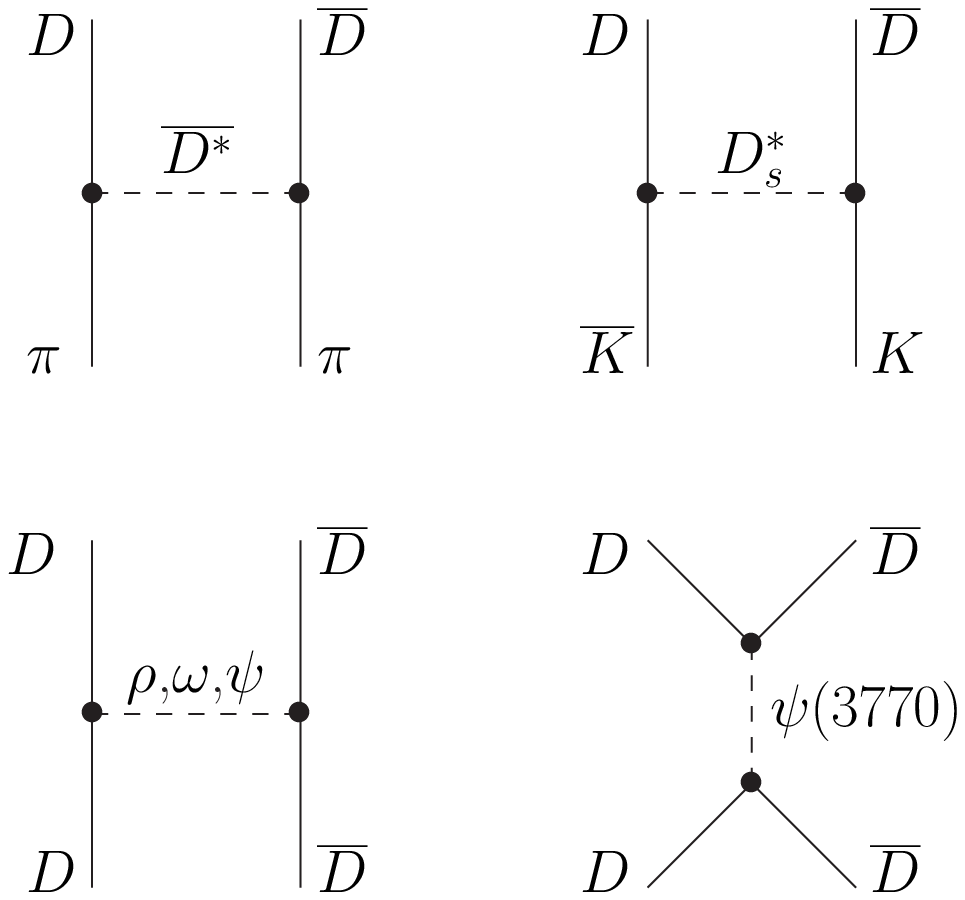,width=11.0cm,height=13.0cm}}
\vspace*{-7.5cm}
\caption{Additional diagrams that arise when the
$\ddbar$ channel is included. 
}
\label{Diag2}
\end{figure}

The interaction in the $\ddbar$ system is constructed 
along the lines of the J\"ulich meson exchange model for the $\pi\pi$ 
interaction whose evaluation has been discussed in detail in 
Refs.~\cite{Lohse90,Janssen95}. The present interaction is based 
on the version described in the latter reference. 
The potentials for $\pi\pi\rightarrow\pi\pi$, $\pi\pi\rightarrow
K\overline{K}$ and $K\overline{K} \rightarrow K \overline{K}$ are
generated from the diagrams shown in Fig.~\ref{Diag1}. 
The additional diagrams that arise for the $\ddbar$ 
potential and for the transitions from $\pi\pi$ and/or
$\kkbar$ to $\ddbar$ are shown in Fig.~\ref{Diag2}. 
In this extension we were guided by SU(4) symmetry \cite{HK14}. 
Thus, we included $t$-channel exchanges of those vector mesons 
which are from the same SU(4) multiplet as those included
in the original J\"ulich model and, moreover, we assumed that
all coupling constants at the additional three-meson vertices 
are given by SU(4) relations.  
Indeed, in Ref.~\cite{HK14} an even more extended model was considered 
which included also the coupling to channels involving the charmed strange 
meson $D_s$(1969). It turned out that the $\ddbar$ production cross sections
based on the $\ddbar$ interactions with or without coupling to $D^+_sD^-_s$ 
are almost identical and, therefore, in the present work we show only the results 
for the latter case. 

The scattering amplitudes are obtained by solving a coupled channel 
scattering equation for these potentials which is formally given by
\begin{eqnarray}
T^{i,j}=&& V^{i,j} + \sum_l V^{i,l} G^l T^{l,j}  \ ,
\label{LS}
\end{eqnarray}
with $i,j,l=\pi\pi, \ \pi\eta, \ \kkbar, \ \ddbar$.

In an approach like ours the $\psi(3770)$ (in the following usually called $\psi$ 
to simplify the notation) has to be included as bare resonance. It aquires its
physical mass and its width when the corresponding potential is iterated 
in the scattering equation (\ref{LS}).
We include the $\psi$ only in the direct $\ddbar$ potential, i.e. in $V^{\ddbar,\ddbar}$. 
The corresponding diagram is depicted on the lower right side of Fig.~\ref{Diag2}.
The potential can be written in the form 
\begin{eqnarray}
V^{\ddbar,\ddbar} = \gamma_0^{\ddbar} \frac{1}{E-m_0} \ \gamma_0^{\ddbar} \ ,
\label{Pole}
\end{eqnarray}
with a bare $\ddbar\psi$ vertex function $\gamma_0^{\ddbar}$ and a bare $\psi(3770)$ 
mass $m_0$. 
Explicit expressions for the vertex function of two pseudoscalar mesons coupled to
a vector meson in the $s$-channel and for the resulting potential can be found in the 
Appendix of Ref.~\cite{Janssen95}.  
The bare mass $m_0$ and the bare coupling constant in $\gamma_0^{\ddbar}$ 
are adjusted in such a way that the resulting $T$-matrix, $T^{\ddbar,\ddbar}$,
has a pole at the physical values of the $\psi(3770)$ resonance. The main results 
shown in the present study are based on a $\ddbar$ model that produces a pole at
$E=(3773 - {\rm i}\,13.6)$ MeV, i.e. at the value specified as ``our fit'' by the 
PDG \cite{PDG}, but we consider also the value obtained in Ref.~\cite{Anashin} for the
mass which is about 6 MeV larger. In the actual calculation the values for 
the mass and width of the $\psi(3770)$ were determined by evaluating the speed plot 
for $T^{\ddbar,\ddbar}$ for simplicity reasons. In addition, we use an isospin averaged 
mass for the $D$ ($\bar D$), namely $1866.9$~MeV. 

\begin{figure}
\includegraphics[height=80mm,angle=-90]{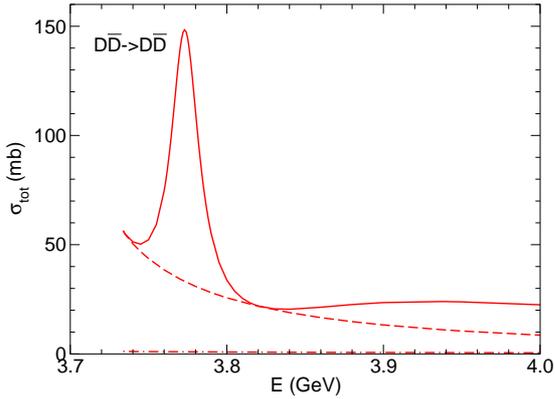}
\caption{Total cross section for $\ddbar$ scattering as a function of the 
center-of-mass energy for isospin $I=0$ (solid line) and $I=1$ (dash-dotted line).
The $s$-wave contribution to the $I=0$ cross section is indicated by the
dashed line. 
}
\label{DDb}    
\end{figure}

Cross sections for $\ddbar$ scattering in the
isospin $I=0$ and $I=1$ states can be found in Fig.~\ref{DDb}.
For the $I=0$ case we show the result from the $s$ wave
separately (dashed curve) so that one can see the impact due to the
$\psi(3770)$ resonance. Note that there is actually a noticeable 
non-resonant contribution in the $p$ wave which is clearly visible 
for energies away from the $\psi(3770)$, i.e. around $3.9$ GeV and above. 
The cross section in the $I=1$ state is very small. It is only in the
order of $1$ mb and, therefore, hardly visible in Fig.~\ref{DDb}.
In this context let us mention that there are other model results on the 
$\ddbar$ interaction in the literature \cite{Liu10}, achieved 
likewise in a meson-exchange approach. 

\begin{figure}
\includegraphics[height=80mm,angle=-90]{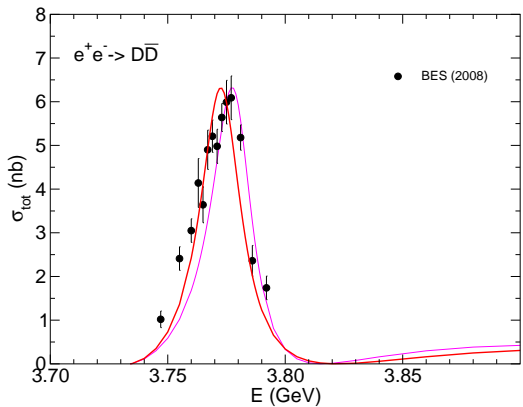}
\caption{
Integrated cross sections for $e^+e^-\to \ddbar$. 
Our results based on the Migdal-Watson approximation are shown by the thick solid line
($m_\psi = 3773$ MeV) and the thin solid line ($m_\psi = 3779$ MeV), respectively. 
Data are taken from Ref.~\cite{Ablikim}. 
}
\label{DDbee}    
\end{figure}

For completeness reasons we present here also results for the 
reaction $e^+ e^- \to \ddbar$, cf. Fig.~\ref{DDbee}. Results corresponding to
a $\psi$ mass of $3773$ or $3779$ MeV, respectively, are indicated by the thick 
and thin solid lines. The pertinent calculation 
was performed in the Migdal-Watson approximation \cite{Watson,Migdal}, i.e. 
by assuming that $T^{\ddbar, \,e^+ e^-} \propto T^{\ddbar, \ddbar}$, 
so that the cross section for $e^+ e^- \to \ddbar$ 
is given by
\begin{equation}
\sigma_{e^+ e^- \to \ddbar} \approx N \ q_{\ddbar} \ |T^{\ddbar,\,\ddbar}|^2 \ .
\end{equation}
Here $q_{\ddbar}$ is the $\ddbar$ center-of-mass momentum and $N$ an
arbitrary normalization factor that has to be adjusted to the data. 
The shown curve is based on the $\ddbar$ $I=0$ amplitude in the $p$ wave.
In this case a factor $q^2_{\ddbar}$ has to be divided out 
because in $e^+ e^- \to \ddbar$ the incoming $\ddbar$ state is not on-shell and, 
therefore, does not feel the angular-momentum threshold factor \cite{Kang15}. 

Our result is in remarkably good agreement with the measured line shape. In particular,
the strong fall-off after the peak value is very well described. As mentioned, we adjusted 
our resonance parameters to the ``our fit'' value of the PDG. No fine tuning to the actual
data was done because we are primarily interested in studying the effect of the $\psi(3770)$
on $\ppbar \to \ddbar$ observables and not on pinning down its resonance parameters.
As already said, there is an uncertainty in the order of $6$ MeV with regard to the actual 
resonance mass \cite{PDG,Anashin}.
For more detailed analyses of the $e^+ e^- \to \ddbar$ data,
see Refs.~\cite{Ablikim,Anashin,Zhang09,Achasov12,Chen12,Liu10}.

\begin{figure*}
\includegraphics[height=80mm,angle=-90]{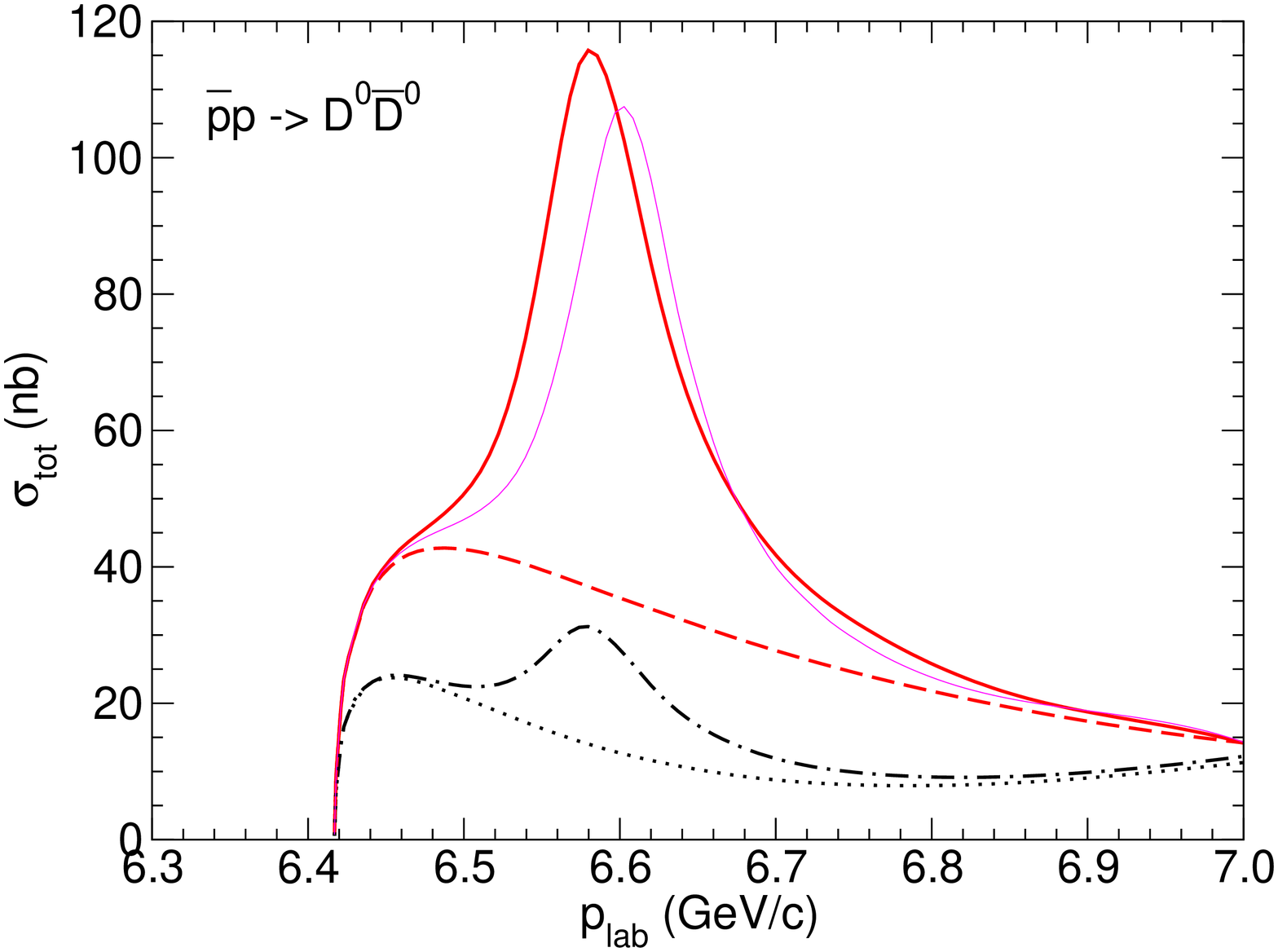}\includegraphics[height=80mm,angle=-90]{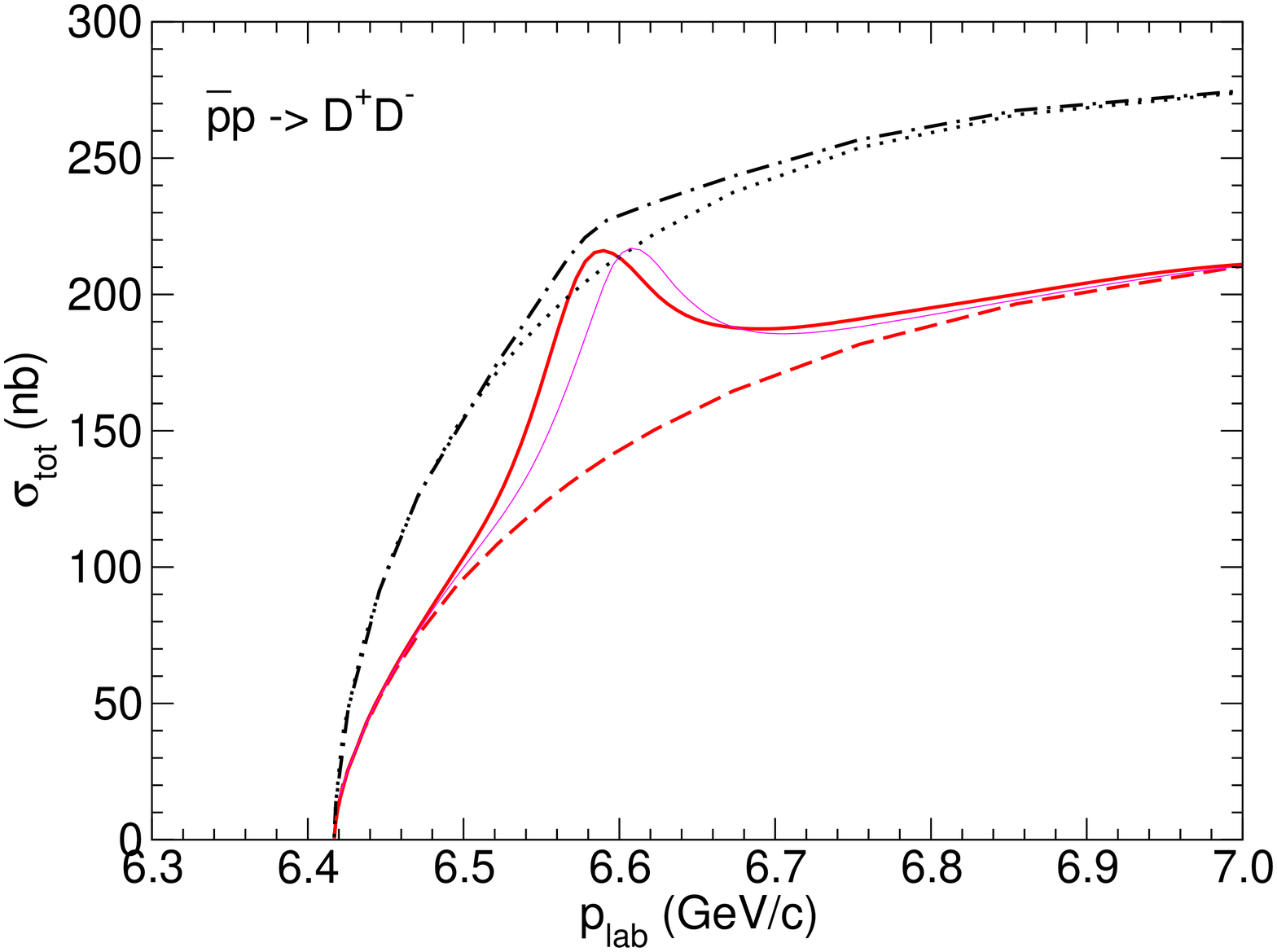}
\caption{Total reaction cross sections for $\pbarp \to \ddbar$ as a function of $p_{lab}$. 
The (red) solid and dashed curves and the (black) dash-dotted and dotted curves
are results for different $\ppbar$ initial-state interactions, see text. 
Solid and dash-dotted curves show the full results with inclusion of the $\psi(3770)$ in 
the FSI, while dashed and dotted curves are without $\psi(3770)$. 
The (magenta) thin solid curves indicate results based on a $\ddbar$ FSI fitted to the higher $\psi$ mass 
($3779$ MeV) obtained in Ref.~\cite{Anashin}. 
}
\label{fig:B} 
\end{figure*}

\begin{figure*}
\includegraphics[height=75mm,angle=-90]{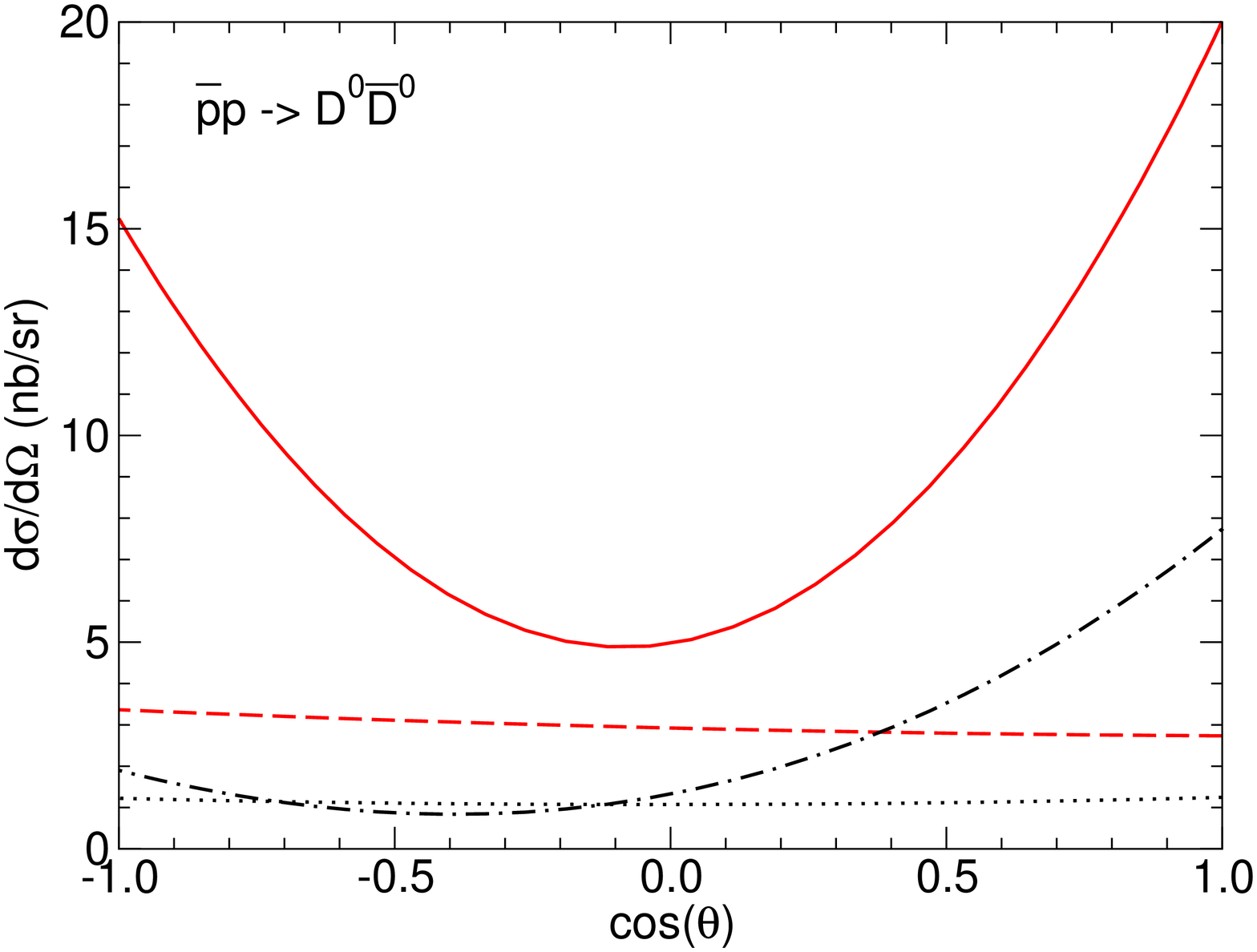}
\includegraphics[height=75mm,angle=-90]{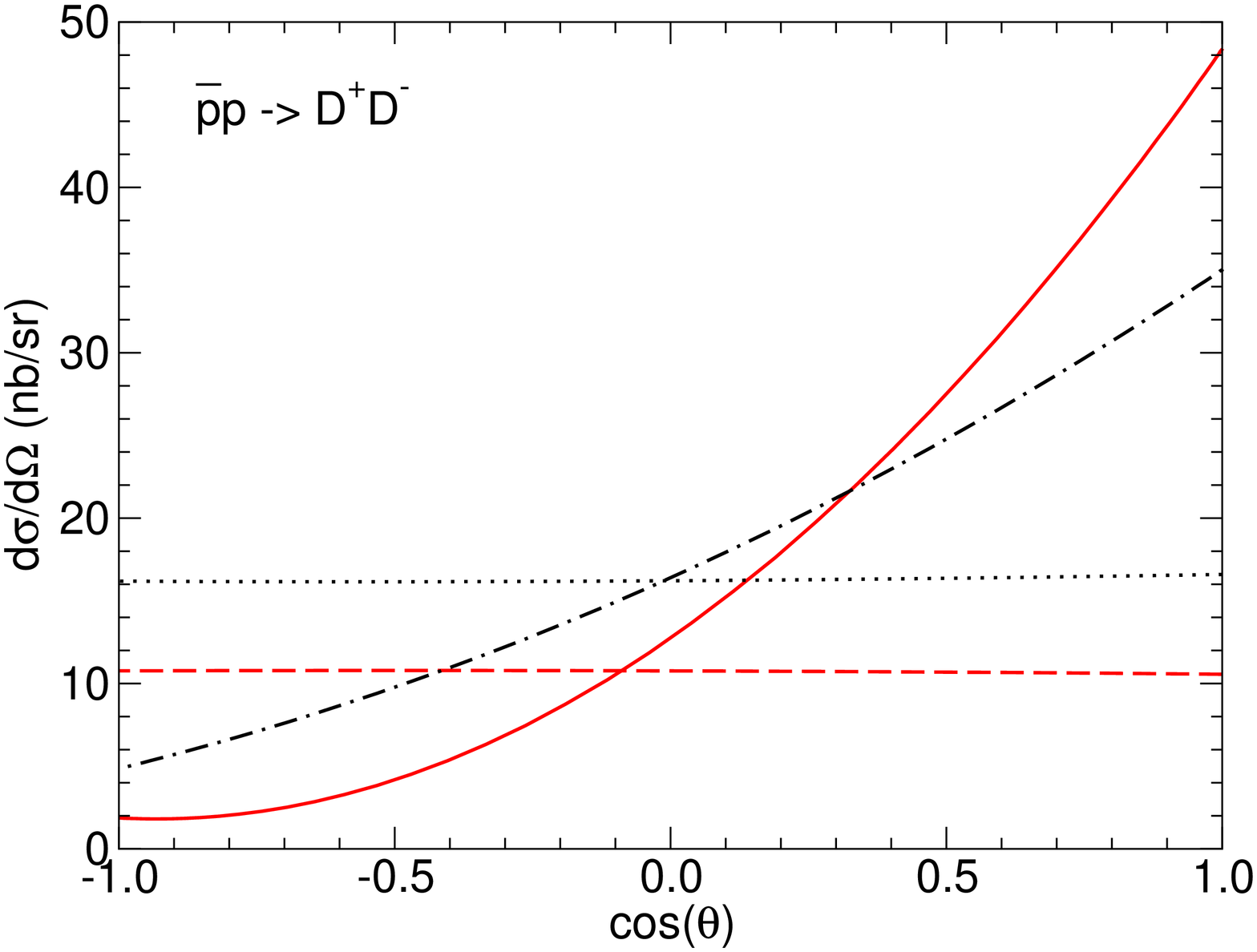}
\caption{Differential cross sections for
$\ppbar \to \ddbar$ at $p_{lab} = 6.578$ GeV/c (excess energy $\epsilon = 40$ MeV).
Same description of curves as in Fig.~\ref{fig:B}.
}
\label{fig:C}  
\end{figure*}

\section{Results for the reaction $\pbarp \to \ddbar$}
\label{sec:3}

Results for the reaction $\pbarp \to \ddbar$ are presented in Figs.~\ref{fig:B} 
and \ref{fig:C}. We performed again calculations
for all $\pbarp$ ISI considered in our previous work \cite{HK14}. However, we
refrain from showing bands here because we wanted to highlight the impact of the
$\psi(3770)$ and this information would have been otherwise hidden in the band. 
Nonetheless, the variations due to the ISI can be still read off because we present curves 
corresponding to the largest and smallest predictions for the $\pbarp \to D^0\bar D^0$
and $\pbarp \to D^+D^-$ cross sections.

The solid and dash-dotted lines in Figs.~\ref{fig:B} and \ref{fig:C} are results with inclusion 
of the $\psi(3770)$ resonance while the dashed and dotted curves are based on 
the $\ddbar$ FSI without the $\psi$, i.e. correspond to the case considered already 
in Ref.~\cite{HK14}. Please recall that in our calculation the $\psi$ is only included 
in the $\ddbar$ interaction and, therefore, its effects come exclusively from the 
corresponding FSI. In principle, there could be also a 
coupling of the (bare) $\psi$ to the $\nnbar$ system which would then contribute to the
$\pbarp \to \ddbar$ transition potential. However, we assume here that this possible 
coupling is negligibly small -- because it is suppressed by the OZI rule. After all, it
requires the annihilation of 3 up- and/or down quark pairs and the creation of a
$c\bar c$ quark pair; see discussion in Sec.~\ref{sec:estimates}. 
 
It is obvious that the $\psi(3770)$ resonance leads to a pronounced enhancement
in the $\ddbar$ production cross section. Especially, for the $D^0\bar D^0$ 
final state the cross section predicted around the resonance peak is now practically
twice as large. Quantitatively, our calculation suggests that the cross section 
due to the $\psi(3770)$ resonance could be the order of $20$ to $80$ nb. 
The effect is less dramatic in the $D^+D^-$ case because here the $s$-wave
contributions are fairly large.
Still for both reactions there should be good chances that one can see a clear 
resonance signal over the background (that comes mostly from the $s$ wave) in
an actual experiment. 

A comparison of the two results with and of the two without $\psi(3770)$ resonance
(solid versus dashed line or dash-dotted versus dotted line, respectively) 
in Fig.~\ref{fig:B} allows one to see the uncertainties in the predictions due to the 
employed $\pbarp$ interactions. Those are substantial but not really dramatic as already 
discussed in Ref.~\cite{HK14}. There is a somewhat larger uncertainty with regard to the
contribution of the $\psi(3770)$ resonance which, however, is not too suprising.
Conservation of total angular momentum and parity implies that the
$\ddbar$ $s$ wave is produced from the $\nnbar$ $^3P_0$ partial wave while
the $\ddbar$ $p$ wave is produced from the $\nnbar$ $^3S_1$-$^3D_1$ partial wave.
Since the $^3S_1$ partial wave is much more sensitive to short-range physics than
higher partial waves there is also a stronger variation in the corresponding amplitudes 
for the various $\nnbar$ potentials considered as ISI. 

The thin solid lines indicate what happens if we employ a $\ddbar$ FSI that is fitted to
the higher value given for the $\psi(3770)$ resonance, i.e. to $3779$ MeV \cite{Anashin,PDG}. 
In this case the contribution of the $\psi$ to the $\pbarp \to \ddbar$ cross 
section is smaller by about 15-20 \% and, of course, the peak is shifted to the higher energy. 

Differential cross sections for $\pbarp \to D^0\bar D^0$ and 
$\pbarp \to D^+D^-$ at $p_{lab}= 6.578$ GeV/c are shown in Fig.~\ref{fig:C}.
This momentum corresponds to a total energy of $3773.8$ MeV so that the results 
display the situation practically at the resonance peak. 
As mentioned before \cite{HK14} without the $\psi(3770)$ the $\ddbar$ pair is
produced predominantly in the $s$ wave, cf. the dashed/dotted curves. Of course, this
changes drastically once the resonance is included (solid/dash-dotted curves). Still, 
there is a strong interference between the $s$- and $p$ waves, especially in the $D^+D^-$ case,
so that the actual angular distribution does not resemble the one one expects from
pure $p$-wave scattering.

\section{Estimates of the cross section for $\ppbar \to \ddbar$}
\label{sec:estimates}

It is possible to provide a rough estimation of the $\ppbar \to \ddbar$ cross
section around the $\psi(3770)$ peak using experimental information by exploiting the 
fact that at energies very close to the resonance the reaction amplitude can be well 
approximated by 
\begin{eqnarray}
T^{i,j} \approx \gamma^{i} \frac{1}{E-m_\psi+{\rm i} \Gamma_\psi / 2} \ \gamma^{j} \ , 
\label{BW}
\end{eqnarray}
an expression which is indeed exact at the $\psi(3770)$ pole.
Here, $\gamma^{i}$ is the dressed vertex function with $i,j= \ddbar$, $\ppbar$,
$e^+e^-$, and $m_\psi$ and $\Gamma_\psi$ the physical mass and width of the
$\psi(3770)$, respectively.

The cross section resulting from this amplitude can be cast into the form
\begin{eqnarray}
\nonumber 
\sigma_{i\to j} \approx \frac{(2J+1)}{(2S^a_i+1)(2S^b_i+1)}\frac{4\pi}{q^2_i} 
\frac{\Gamma_{i}/2 \ \Gamma_{j}/2}{(E-m_\psi)^2+\Gamma^2_\psi/4} \ , \\
\label{BWC}
\end{eqnarray}
i.e. into the standard Breit-Wigner formula. Here $J$ is the total 
angular momentum, $S^a_i$ and $S^b_i$ are the spins of the particles in the
incoming channel, $q_i$ is the center-of-mass momentum in the incoming
channel, and $\Gamma_{i}$ and $\Gamma_{j}$ are the partial widths for the
decay of the $\psi (3770)$ into the channels $i$ and $j$.

There are cross section data on the reactions $\eebar \to \ddbar$ \cite{Ablikim}
and $\eebar \to \ppbar$ \cite{Ablikim2} in the vicinity of the $\psi(3770)$
resonance. Together with unitarity constraints for $\ddbar \to \ddbar$ 
they allow one to determine all the $\Gamma_{i}$'s needed for
estimating the $\ppbar \to \ddbar$ cross section.
With regard to the $\ddbar \to \ddbar$ cross section, we assume for simplicity that 
$\Gamma_{\ddbar}\approx \Gamma_{\psi}$ -- which is anyway consistent with the PDG listing
\cite{PDG} -- so that the cross section is given by the unitarity bound. 
At the energy corresponding to the $\psi(3770)$ resonance this amounts to 
$\sigma_{\ddbar \to \ddbar} \approx 200$ mb.
Assuming that the $I=1$ $p$-wave amplitude is small, which is indeed the case 
in our model (cf. Fig.~\ref{DDb}) yields $\sigma_{\ddbarc\to \ddbarc} 
\approx 50$ mb.
The measured $\eebar \to \ddbarc$ cross section at the $\psi(3770)$ resonance 
peaks at about $3$ nb \cite{Ablikim} while the analysis of 
the $\eebar \to \ppbar$ cross section performed in Ref.~\cite{Ablikim2}
suggests that the contribution due to the $\psi(3770)$ resonance could amount
to either $0.059^{+0.070}_{-0.020}$ or $2.57^{+0.12}_{-0.13}$ pb. Combining those 
results and using Eq.~(\ref{BWC}) to determine the various $\Gamma_{i}$'s we deduce for 
$\ppbar \to \ddbarc$ a cross section of either $3\sim 18$ nb or around $350$ nb. 
There is a large uncertainty for the lower value that follows from the BESIII analysis
but, still, one could argue that it is roughly in line with our theory prediction.

As alternative let us consider also an estimate that follows from a QCD-based perturbative 
approach to charmonium decays into baryon-antibaryon pairs \cite{Bolz,Brambilla}. 
That approach relies on a factorized expression for the $\psi\to\ppbar$ decay amplitude into 
a $c\bar c \to \ppbar$ annihilation amplitude and the $\psi$ wave function at the 
origin -- see, e.g., Eqs.~(4.68)--(4.71) of Ref.~\cite{Brambilla}. The model dependence regarding 
the $c\bar c \to \ppbar$ annihilation amplitude, which involves at least three gluons in
the creation of up and down quark-antiquark pairs, can be eliminated considering the ratio
of $\Gamma_{\psi \rightarrow p \bar p}$ to $\Gamma_{J/\psi \rightarrow p \bar p}$. 
If, in addition, one assumes that the strong as well as the electronic decays of $\psi(3770)$ 
proceed predominantly through the $2\,^3S_1$ component of its wave function (recall that the existence 
of an important $2\,^3S_1$ component comes from the large decay width of $\psi(3770)$ 
into $e^+e^-$), one can express $\Gamma_{\psi \rightarrow \pbarp}$ in terms of measured 
quantities:
\bea
\nonumber
\Gamma_{\psi \rightarrow \pbarp} = 
\left(\frac{1 - 4m^2_p/M^2_{\psi}}{1 - 4m^2_p/M^2_{J/\Psi}}
\right)^{\frac{1}{2}} \frac{\Gamma_{\psi \rightarrow e^+e^-}} 
{\Gamma_{J/\Psi \rightarrow e^+e^-}} \ 
{\Gamma_{J/\Psi \rightarrow \pbarp}} \ . \\
\label{Gpsi-fin}
\eea
Using PDG values~\cite{PDG} for the electronic decay widths and for the width of 
$J/\Psi \rightarrow \pbarp$, one obtains from Eq.~(\ref{BWC}) for the cross
section $\sigma_{\ppbar \to \ddbarc} \approx 0.2\,{\rm nb}$, due to the $\psi$(3770),
which is significantly lower than the value extracted from the BESIII result 
but also much smaller than our model predictions.  

\section{Summary}
\label{sec:4}

We have presented a study of the reaction $\pbarp \to \ddbar$ close to the 
production threshold with special emphasis on the role played by the 
$\psi$(3770) resonance. 
The work is an extension of a recent calculation by us \cite{HK14}
which is performed within a meson-baryon model where the elementary 
charm production process is described by baryon exchange, and where
effects of the interactions in the initial ($\ppbar$) and final ($\ddbar$) 
states are taken into account rigorously. 
The $\psi$(3770) resonance, which is included in the $\ddbar$
FSI in the present calculation, produces a sizeable enhancement in the
$\pbarp \to \ddbar$ cross section around the resonance energy. 
Indeed, our predictions for the total $\ddbar$ production cross section in
the considered near-threshold region are in the order of $30$ -- $250$ nb, 
where the contribution from the $\psi$(3770) resonance itself amounts to 
roughly $20$ -- $80$ nb.
Given the magnitude and the shape of the cross section due to the $\psi$(3770)
there should be good chances to measure the pertinent contribution in 
dedicated experiments which could be performed at FAIR. 

\section*{Acknowledgements}
Work partially supported by the Brazilian agencies Conselho Nacional 
de Desenvolvimento  Cient\'{\i}fico e Tecnol\'ogico - CNPq, Grant No. 305894/2009-9, 
and Funda\c{c}\~ao de Amparo \`a Pesquisa do Estado de S\~ao Paulo - FAPESP, Grant No. 
2013/01907-0.



\begin{thebibliography}{20}
%
\bibitem{HK14}
  J. Haidenbauer and G. Krein, 
  Phys.\ Rev.\ D {\bf 89}, 114003 (2014). 

\bibitem{Goritschnig13} 
  A.~T.~Goritschnig, B.~Pire, and W.~Schweiger,
  Phys.\ Rev.\ D {\bf 87}, 014017 (2013); Phys.\ Rev.\ D {\bf 88}, 079903(E) (2013). 

\bibitem{Mannel12} 
  A.~Khodjamirian, C.~Klein, T.~Mannel, and Y.~M.~Wang,
  Eur.\ Phys.\ J.\ A {\bf 48}, 31 (2012).

\bibitem{Titov:2008yf}
  A.~I.~Titov and B.~K{\"a}mpfer,
  Phys.\ Rev.\  C {\bf 78}, 025201 (2008).

\bibitem{Kerbikov}
  B. Kerbikov and D. Kharzeev,
  Phys.\ Rev.\  D {\bf 51}, 6103 (1995).

\bibitem{Kaidalov:1994mda}
  A.~B.~Kaidalov and P.~E.~Volkovitsky,
  Z.\ Phys.\  C {\bf 63}, 517 (1994).

\bibitem{Kroll:1988cd}
  P.~Kroll, B.~Quadder and W.~Schweiger,
  Nucl.\ Phys.\  B {\bf 316}, 373 (1989).

\bibitem{Ablikim} M. Ablikim et al., Phys. Lett. B {\bf 668}, 263 (2008).

\bibitem{Anashin} V.V. Anashin et al., Phys. Lett. B {\bf 711}, 292 (2012).

\bibitem{PDG} K.A. Olive et al. (Particle Data Group), 
   Chin. Phys. C {\bf 38}, 090001 (2014). 

\bibitem{PANDA} W.~Erni et al., arXiv:0903.3905 [hep-ex].
\bibitem{Wiedner} U.~Wiedner,
  Prog.\ Part.\ Nucl.\ Phys.\  {\bf 66}, 477 (2011).
\bibitem{Prencipe2014} E.~Prencipe,
  arXiv:1410.5680 [hep-ex].


\bibitem{Mull}
  V.~Mull and K.~Holinde,
  Phys.\ Rev.\  C {\bf 51}, 2360 (1995)

\bibitem{Haidenbauer2007}
  J.~Haidenbauer, G.~Krein, U.-G.~Mei{\ss}ner, and A.~Sibirtsev,
  Eur.\ Phys.\ J.\  A {\bf 33}, 107 (2007).
\bibitem{Haidenbauer2008}
  J.~Haidenbauer, G.~Krein, U.-G.~Mei{\ss}ner, and A.~Sibirtsev,
  Eur.\ Phys.\ J.\  A {\bf 37}, 55 (2008).

\bibitem{Hai10}
  J.~Haidenbauer, G.~Krein, U.-G.~Mei{\ss}ner, and L.~Tolos,
  Eur.\ Phys.\ J.\ A {\bf 47}, 18 (2011).

\bibitem{HK10}
  J. Haidenbauer and G. Krein, Phys. Lett. B {\bf 687}, 314 (2010). 
 
\bibitem{Haidenbauer:1991kt}
  J.~Haidenbauer, T.~Hippchen, K.~Holinde, B.~Holzenkamp, V.~Mull, and J.~Speth,
  Phys.\ Rev.\  C {\bf 45}, 931 (1992).
  
\bibitem{Kohno} M. Kohno and W. Weise, Phys. Lett. B {\bf 179}, 15 (1986).

\bibitem{Lohse90} D. Lohse, J.~W. Durso, K. Holinde, and J. Speth,
Nucl. Phys. {\bf A516}, 513 (1990).

\bibitem{Janssen95} G. Janssen, B.C. Pearce, K. Holinde, and J. Speth,
Phys. Rev. D {\bf 52}, 2690 (1995).

\bibitem{Liu10} 
  Y.-R.~Liu, M.~Oka, M.~Takizawa, X.~Liu, W.~-Z.~Deng, and S.-L.~Zhu,
  Phys.\ Rev.\ D {\bf 82}, 014011 (2010).

\bibitem{Watson}
        K.M. Watson, Phys. Rev. {\bf 88}, 1163 (1952).
\bibitem{Migdal}
        A.B. Migdal, JETP {\bf 1}, 2 (1955).
\bibitem{Kang15} 
  X.~W.~Kang, J.~Haidenbauer and U.-G.~Mei\ss ner,
  Phys.\ Rev.\ D {\bf 91}, 074003 (2015).
 
\bibitem{Zhang09} 
  Y.-J.~Zhang and Q.~Zhao,
  Phys.\ Rev.\ D {\bf 81}, 034011 (2010).
\bibitem{Achasov12} 
  N.~N.~Achasov and G.~N.~Shestakov,
  Phys.\ Rev.\ D {\bf 86}, 114013 (2012). 
\bibitem{Chen12} 
  G.-Y.~Chen and Q.~Zhao,
  Phys.\ Lett.\ B {\bf 718}, 1369 (2013).
\bibitem{Achasov13} 
  N.~N.~Achasov and G.~N.~Shestakov,
  Phys. Rev. D {\bf 87}, 057502 (2013).

\bibitem{Ablikim2} M.~Ablikim {\it et al.}, 
  Phys.\ Lett.\ B {\bf 735}, 101 (2014). 

\bibitem{Bolz} 
  J.~Bolz and P.~Kroll,
  Eur.\ Phys.\ J.\ C {\bf 2}, 545 (1998).

\bibitem{Brambilla} N. Brambilla et al., arXiv:hep-ph/0412158.

\end{thebibliography}
\end{document}